\documentclass[11pt]{svproc}
\usepackage{amsmath,amssymb,times}
\usepackage{graphicx,colordvi,color}

\usepackage{url}

\newcommand{\sech}{\,{\sf sech}}

\begin{document}

\mainmatter              
\title{Signatures of chaotic dynamics in wave motion according to the extended KdV equation}
\titlerunning{Signatures of chaotic dynamics}
\author{Anna Karczewska\inst{1} \and Piotr Rozmej\inst{2}}
\authorrunning{Anna Karczewska and Piotr Rozmej}
\institute{Faculty of Mathematics, Computer Science and Econometrics,\\ University of Zielona G\'ora, 65-246 Zielona G\'ora, Poland,\\
\email{A.Karczewska@wmie.uz.zgora.pl},\\ WWW home page:
\texttt{http://staff.uz.zgora.pl/akarczew}
\and
Faculty of Physics and Astronomy, University of Zielona G\'ora,\\ 65-246 Zielona G\'ora, Poland,\\
\email{P.Rozmej@if.uz.zgora.pl}}

\maketitle              

\begin{abstract} 
In this communication we test the hypothesis that for some initial conditions the time evolution of surface waves according to the \emph{extended KdV equation} (KdV2)
exhibits signatures of the deterministic chaos.
\end{abstract}


\section{Introduction} \label{intro}

Korteweg-de Vries equation (KdV) is the most famous nonlinear partial differential equation modelling long-wave, weekly dispersive gravity waves of small amplitude on a surface of shallow water. In scaled variables and a fixed reference frame KdV takes the following form
\begin{equation} \label{kdv}	
\eta_t+\eta_x +\frac{3}{2}\alpha \eta\eta_x+\frac{1}{6}\beta\eta_{xxx} =0.
\end{equation}
In (\ref{kdv}), $\eta(x,t)$ represents the wave profile, $\alpha\! =\! A/H$, $\beta\!= \!(H/L)^{2}$, where $A$ is wave's amplitude, $L$ - it's average wavelength and $H$ is water depth. Indexes indicate partial derivatives. KdV equation (\ref{kdv}) is derived from Euler equations in perturbation approach, under assumption that parameters $\alpha\!\approx\!\beta$ are small. 
It has several analytic solutions: single soliton, multi-soliton, and periodic (cnoidal) ones. KdV is integrable. It also has the unique property of the infinite number of integral invariants, see, e.g., \cite{DrJ}.

In 1990, Marchant and Smyth, extending perturbation approach to second order in small parameters $\alpha,\beta$, derived the \emph{extended KdV equation} (KdV2) \cite{MS90}
\begin{align} \label{nkdv2} \eta_t+\eta_x & +\frac{3}{2}\alpha \eta\eta_x+\frac{1}{6}\beta\eta_{xxx} -\frac{3}{8}\alpha^2\eta^2\eta_x \\ & + \alpha\beta \left(\frac{23}{24}\eta_x\eta_{xx} +\frac{5}{12}\eta\eta_{xxx} \right) + \frac{19}{360}\beta^2\eta_{xxxxx}=0. \nonumber \end{align}
Studying this equation, we showed that KdV2 has only one exact invariant, representing mass (volume) of displaced fluid. The other integral invariants are only approximate, with deviations of the order of $O(\alpha^{2})$ \cite{KRIR}. Next, we showed that KdV2 equation, despite being non-integrable, has exact single soliton and periodic solutions in the same form as KdV equation, but with slightly different coefficients \cite{KRI,IKRR,RKI}.

An exact single soliton solution of KdV2 has the same form as the KdV soliton, that is, 
\begin{equation} \label{kdv2sol}
\eta(x,t)= A\, \sech [B(x-vt)].
\end{equation}
However, coefficients $A, B, v$ are uniquely determined by the coefficients of the KdV2 equation \cite{KRI}. This property is entirely different from KdV solitons' properties, for which there is a one-parameter family of possible solutions. Therefore KdV equation admits multi-soliton solutions, whereas the KdV2 equation does not. 
Since the equation (\ref{nkdv2}) is second-order in small parameters $\alpha,\beta$ (assuming that $\alpha \approx \beta$), it should be a good approximation for much larger values of small parameters than KdV.

\begin{figure}[ht] \begin{center} 
\resizebox{0.8\columnwidth}{!}{\includegraphics{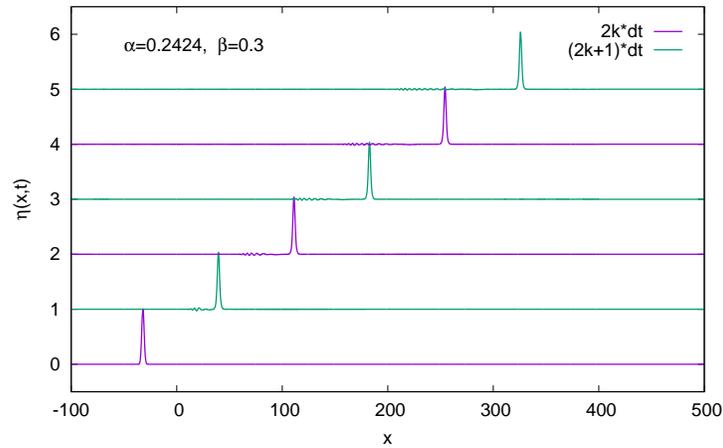}} 
\caption{Snapshots of time evolution according to the KdV2 equation (\ref{nkdv2}). Subsequent profiles corresponding to times $t_{n}=n*64$, with $n=0,1,\ldots,5$, are shifted up by one unit to avoid overlaps. Initial condition in the form of the Gaussian of the same volume as the soliton, the same amplitude and velocity.}
\label{kdv2G1.}
\end{center} 
\end{figure}

\begin{figure}[ht] \begin{center} 
\resizebox{0.8\columnwidth}{!}{\includegraphics{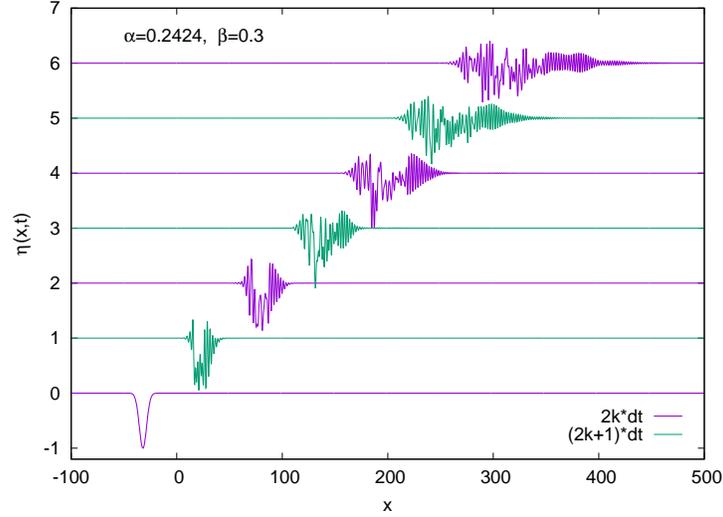}} 
\caption{The same as in Fig.~\ref{kdv2G1.} but for 
initial condition in the form of the Gaussian which volume is three times greater, with the same amplitude and velocity but the inverted displacement.} \label{Ks3i_a24b3d0.}
\end{center} 
\end{figure}

However, in the KdV2 case, for some initial conditions we obtained unexpected results.
In particular, when the initial condition was chosen in the form of depression (instead of `usual' elevation), the time evolution calculated according to the KdV2 equation (\ref{nkdv2}) appeared to be entirely different than that when initial conditions are not much different from the exact soliton. We first encountered these facts in \cite{KRcmst}.
We show an example of this behavior in Figs.\ \ref{kdv2G1.} and~\ref{Ks3i_a24b3d0.}. 

At the first glance, time evolution presented in Fig.~\ref{Ks3i_a24b3d0.} (bottom) looks \emph{chaotic}. In the next section, we try to verify this observation quantitatively.

\section{Dynamics determined by the KdV2 equation} \label{ChaMo}

It is well known \cite{ASY} that the \emph{deterministic chaos} occurs when trajectories of the system's motion, starting from very close initial conditions, diverge exponentially with time. How can we define the distance between the trajectories? Let $\eta_{1}(x,t)$ and $\eta_{2}(x,t)$ denote two different trajectories. 
Define the following measures of the distance between them:
\begin{align} \label{miara1}
M_{1}(t) & = \int_{-\infty}^{\infty} |\eta_{1}(x,t)-\eta_{2}(x,t)|\,dx \\ 
 \label{miara2}
M_{2}(t) & = \int_{-\infty}^{\infty} \left[\eta_{1}(x,t)-\eta_{2}(x,t)\right]^{2} dx .
\end{align}

In numerical simulations, we utilize the finite difference method (FDM) described in detail in \cite{KRI,KRcmst} with $N$ of the order 5000-10000. Integrals are approximated by sums, so $M_{1}(t)\approx \sum_{i=1}^{N} |\eta_{1}(x_{i},t)-\eta_{2}(x_{i},t)|dx$ and $M_{2}(t)\approx \sum_{i=1}^{N} [\eta_{1}(x_{i},t)-\eta_{2}(x_{i},t)]^{2}dx$. 
In calculations, we use the periodic boundary conditions. Therefore, the interval $x\in[x_{1},x_{2}]$ has to be much wider than the region where the surface wave is localized. 
In numerical calculations, presented in \cite{KRcmst}, when the initial conditions were exact KdV2 solitons, the invariant $I_{1}=\int_{-\infty}^{\infty} \eta(x,t) dx$ was conserved with the precision $10^{-10}-10^{-12}$. In calculations shown in this note, since initial conditions are much different, $I_{1}$ is conserved up to $10^{-7}$.

\begin{figure}[ht] \begin{center}
\resizebox{0.8\columnwidth}{!}{\includegraphics{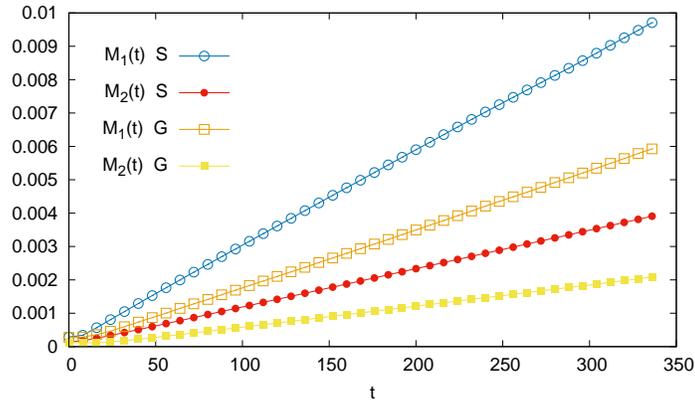}} 
\caption{Distances between trajectories, which start from almost identical initial conditions, as functions of time. Lines marked by S correspond to initial conditions in the form of exact KdV2 soliton (\ref{kdv2sol}), whereas those marked by G correspond to initical conditions in the form of Gaussians. Open symbols indicate $M_{1}$ measures, whereas the filled ones indicate $M_{2}$ measures. } \label{kdv2SolG1.} \end{center}
\end{figure}

First, we check the divergence of trajectories when the initial conditions are close to the exact KdV2 soliton. This is presented in Fig.~\ref{kdv2SolG1.} with lines marked by~S. The lines marked by~G in Fig.~\ref{kdv2SolG1.} represent the divergence of trajectories when $\eta_{1}(x,t=0)$ was the Gaussian distortion having the same amplitude $A_{1}$ as the KdV2 soliton, with the width $\sigma_{1}$ providing the same volume. Then, for $\eta_{2}(x,t=0)$ we chose the similar Gaussian form but with parameters slightly changed, namely with $A_{2}=A_{1}(1+\varepsilon)$ and $\sigma_{2}=\sigma_{1}/(1+\varepsilon)$, which ensures the same initial volume. In both cases, initial velocity was assumed to be equal to soliton's velocity. Note, that profiles shown in Fig.~\ref{Ks3i_a24b3d0.} represent the evolution of $\eta_{1}(x,t=0)$.
Results displayed in Fig.~\ref{kdv2SolG1.} were obtained for $\varepsilon=10^{-4}$. 
For $\varepsilon=10^{-3}$ and $\varepsilon=10^{-5}$ the relative distances $M(t)/M(0)$
behave almost exactly the same. 
It is clear that in all cases presented in Fig.~\ref{kdv2SolG1.} the distances between trajectories which start from neighbor initial conditions diverge linearly with time. For dynamical systems, it means that such conditions belong to this phase-space region in which motion is regular. 

\begin{figure}[htb] \begin{center} \resizebox{0.8\columnwidth}{!}{\includegraphics{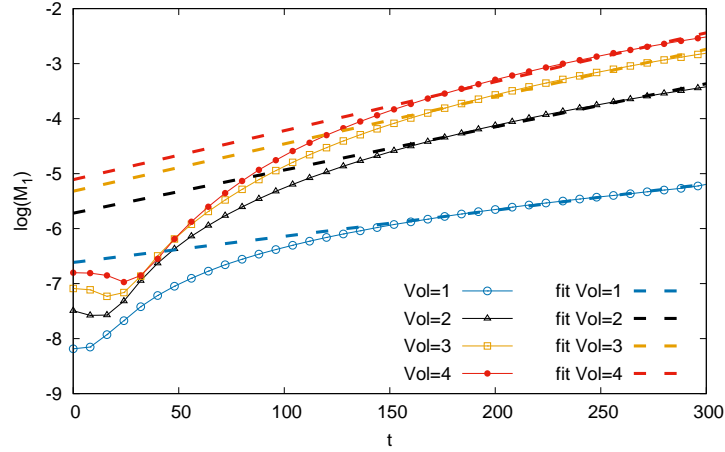}} \caption{Time dependence of measures $M_{1}$ 
for $\varepsilon=10^{-4}$ and different volumes of initial conditions.} \label{M_1V1-4Long.} \end{center} \end{figure}

In the below-presented calculations, we studied the motion according to the KdV2 equation when initial displacements have the sign opposite from the soliton. For $\eta_{1}(x,t=0)$ we chose the Gaussian distortion having the amplitude $-A_{1}$ of the KdV2 soliton, but with the width $\sigma_{1}$ ensuring multiple soliton's volumes. For $\eta_{2}(x,t=0)$ we chose the similar Gaussian form but with parameters slightly changed, namely with $A_{2}=-A_{1}(1+\varepsilon)$ and $\sigma_{2}=\sigma_{1}/(1+\varepsilon)$, which ensures the same initial volume. In both cases, the initial velocity is chosen the same as the velocity of the KdV2 soliton.

Since in all cases shown below, the distances between trajectories increased much faster than linearly with time, the next figures are plotted on a semilogarithmic scale. 
In Fig.~\ref{M_1V1-4Long.}, we show the time dependence of the distance measures $M_{1}$ for $\varepsilon=10^{-4}$. The notation Vol=$n$, with $n=1,2,3,4$ denotes the initial volume of the $\eta_{1}(x,t=0)$ in the units of the KdV2 soliton volume.
In Fig.\ 5 
we present the analogous results but for $M_{2}$ measures. In both figures we observe almost perfect exponential divergence of trajectories, even for $n=1$. Fits to the plots displayed in Fig.~\ref{M_1V1-4Long.}
give the following exponents: 0.00475, 0.00785, 0.0086 and 0.0089 for Vol=1,2,3,4, respectively. For $M_{2}$ measures the fitted exponents are: 0.00477, 0.0051, 0.00539 and 0.00561, respectively. All these exponents are obtained by fitting in the interval $t\in[150:300]$. 

\begin{figure}[hbt] \begin{center} \resizebox{0.8\columnwidth}{!}{\includegraphics{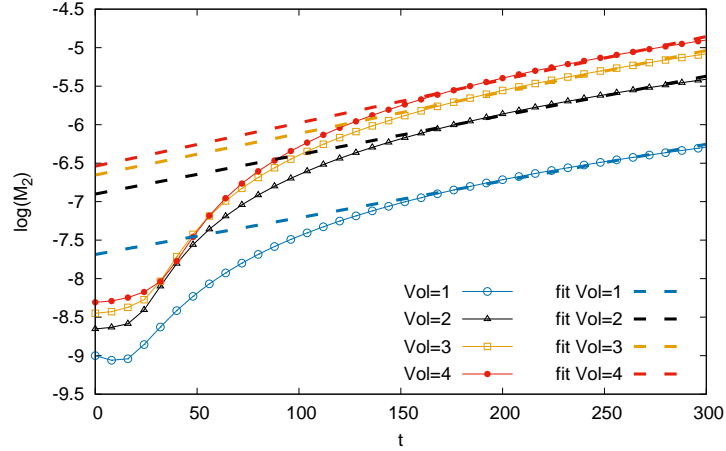}} \caption{The same as in Fig.~\ref{M_1V1-4Long.} but for $M_{2}$ measures. }  \end{center}\label{M_2V1-4Long.} \end{figure}

The results shown above allow us to conclude that for initial conditions substantially different from the exact KdV2 soliton, the dynamics of motion governed by the KdV2 equation exhibits exponential divergence of trajectories.


\end{document}